\documentstyle[aps,preprint]{revtex} 
\newcommand\Tr{\mbox{\rm Tr}}
\newcommand{\beq}{\begin{equation}}
\newcommand{\eeq}{\end{equation}}
\newcommand{\half}{\mbox{$\textstyle \frac{1}{2}$} }
\newcommand{\ket}[1]{\left | \, #1 \right \rangle}
\newcommand{\mod}{\mbox{\rm mod}}
\newcommand{\bra}[1]{\left \langle #1 \, \right |}
\newcommand{\proj}[1]{\ket{#1}\!\!\bra{#1}}
\renewcommand{\choose}[2]{{{#1}\atopwithdelims(){#2}}}

\title{Concentrating Partial Entanglement \\ by Local Operations}
\author{Charles H. Bennett \\
{\protect\small\sl IBM Research Division, T.J. Watson Center,
Yorktown Heights, NY 10598, USA.} \\
Herbert J. Bernstein \\
{\protect\small\sl Hampshire College, Institute for Science and
Interdisciplinary Studies,
Amherst, MA, 01002 USA.} \\
Sandu Popescu \\
{\protect\small\sl
Physics Dept. Tel Aviv University, Israel.} \\
Benjamin Schumacher \\
{\protect\small\sl Physics Department, Kenyon College, Gambier,
OH 43022, USA.}}
 
\date{\today}
 
\begin{document}
\maketitle
 
\thispagestyle{empty}
 
\subsection*{\centering Abstract}
{If two separated observers are supplied with entanglement, in the form
of $n$ pairs of particles in identical partly-entangled pure states,
one member of each pair being given to each observer; they can, by
local actions of each observer, concentrate this entanglement into
a smaller number of maximally-entangled pairs of particles,
for example Einstein-Podolsky-Rosen singlets, similarly shared between
the two observers.  The concentration process asymptotically
conserves {\em entropy of entanglement}---the von Neumann entropy of
the partial density matrix seen by either observer---with the yield of
singlets approaching, for large $n$, the base-2 entropy of
entanglement of the initial partly-entangled pure state.  Conversely,
any pure or mixed entangled state of two systems can be produced by
two classically-communicating separated observers, drawing on a
supply of singlets as their sole source of entanglement.}
 
\vfill
 
PACS numbers: 03.65.Bz, 42.50.Dv, 89.70.+c\vfill
 
\section{Introduction}
 
Recent results in quantum information theory have shed light on
the channel resources needed for faithful transmission of
quantum states, and the extent to which these resources can be
substituted for one another.
The fundamental unit of quantum information transmission is
the quantum bit or {\em qubit\/}~\cite{Schu95}.
A qubit is any two-state
quantum system, such as a spin-1/2 particle or an arbitrary superposition of
two Fock states.  If two orthogonal states of the system are used to represent
the classical Boolean values 0 and 1, then a qubit differs from a bit in that it
can also exist in arbitrary complex superpositions of 0 and 1, and it can be
entangled with other qubits.  Schumacher's quantum data compression
theorem~\cite{Schu95,JS94} characterizes the number of
qubits, sent through the channel from sender to
receiver, that are asymptotically necessary and
sufficient for faithfully transmitting unknown pure states
drawn from an arbitrary known source ensemble.

Quantum superdense coding~\cite{BW92} and
quantum teleportation~\cite{BBCJPW93} consume a different quantum
resource---namely entanglement, in the form of maximally
entangled pairs of particles initially shared between sender and
receiver---and use it to assist, respectively, in the performance
of faithful classical and quantum communication.  Following
Schumacher's terminology, we define an {\em ebit\/} as the amount
of entanglement between a maximally entangled pair of two-state
systems, such as two spin-$\half$ particles in the singlet state,
and we inquire how many ebits are needed for various tasks.
In \cite{BBCJPW93}, for example, it is shown that the consumption
of one shared ebit, together with the transmission of a two-bit
classical message, can be substituted for the transmission of one
qubit.
 
An important concept in quantum data transmission is {\sl fidelity\/},
the probability that a channel output would pass a test for being the
same as the input conducted by someone who knows what the input was.
If a pure state $\psi$ sent into a quantum channel emerges as the
(in general) mixed state represented by density matrix $W$, the fidelity
of transmission is defined
as $F=\bra{\psi}W\ket{\psi}$.  A quantum channel will be
considered faithful if in an appropriate limit the expected fidelity
of transmission tends to unity.  This means that the outputs are almost
always either identical to the inputs, or else so close that the
chance of distinguishing them from the inputs by any quantum meausrement
tends to zero.
 
Note that qubits are a directed channel resource, sent in a
particular direction from sender to the receiver; by contrast,
ebits are an undirected resource {\em shared \/} between sender
and receiver.  For example, if you prepare two particles in a
singlet state and give me one of them, the result is the same as
if I had prepared the particles and given you one of them.  Ebits
are a weaker resource than qubits, in the sense that transmission
of one qubit can, as just described, be used to create one ebit
of entanglement; but the sharing of an ebit, or many ebits, does
not by itself suffice to transmit an arbitrary state of a 2-state
quantum system, or qubit, in either direction.  To do that, the
ebits must be supplemented by directed classical bits, as in
teleportation.
 
One would naturally like to know whether, in order to be useful
for purposes such as teleportation, entanglement must be supplied
in the form of maximally entangled pairs.  In particular, could
partly-entangled pure states, such as pairs of particles in the
state
\beq
\cos \theta \ket{\uparrow_A}\otimes\ket{\downarrow_B}
-\sin\theta\ket{\downarrow_A}\otimes\ket{\uparrow_B}
\eeq
be used instead, and, if so, how many such pairs
would be needed to substitute for one maximally entangled pair?
Note that by using the Schmidt decomposition, and absorbing
phases into the definitions of the basis states, any entangled
state can be represented by a bi-orthogonal expression of this
form, with positive real coefficients~\cite{Asher}
\beq
\Psi(A,B)= \sum_{i=1}^d c_i\ket{\alpha_i}\otimes\ket{\beta_i},
\eeq
where $\ket{\alpha_1},\ket{\alpha_2}...\ket{\alpha_d}$ and
$\ket{\beta_1},\ket{\beta_2}...\ket{\beta_d}$
are orthonormal states of subsystems $A$ and $B$ respectively, and the
coefficients $c_i$ are real and positive.  From the viewpoint of either
observer, an entangled state appears as a mixed state, described by
a density matrix obtained by tracing over the degrees of freedom of
the other observer.  These density matrices are diagonal in the
Schmidt basis:
\beq
\rho_A= \Tr_B\proj{\Psi(A,B)} = \sum_i c_i^2\proj{\alpha_i},
\eeq
and similarly for $\rho_B$.
 
The entanglement of a partly-entangled pure state can be
naturally parameterized by its entropy of entanglement, defined
as the von Neumann entropy of either $\rho_A$ or $\rho_B$, or
equivalently as the Shannon entropy of the squares of the
Schmidt coefficients.
\beq
E=-\Tr\rho_A\log_2\rho_A=-\Tr\rho_B\log_2\rho_B =-\sum_ic_i^2\log_2c_i^2.
\eeq
Without losss of generality we choose the $\alpha$ and $\beta$ bases such
that the sequence of Schmidt coefficients $c_1, c_2...$ is nonincreasing.
 
The quantity $E$, which we shall henceforth often call simply
``entanglement,'' ranges from zero for a product state (eg $\theta=0$)
to 1 ebit for a maximally entangled pair of two-state particles (eg
$\theta=\pi/4$).  (More generally, a maximally entangled state
of two subsystems has $d$ equally weighted terms in its Schmidt
decomposition, giving $\log_2d$ ebits of entanglement, where $d$ is
the Hilbert space dimension of the smaller subsystem.)
 
If a partly entangled pair, with $E < 1$, is used directly for
teleportation, unfaithful transmission will result.  If it is
used for superdense coding, the resulting classical channel will
be noisy.  In this paper we show how, by local operations on a large
number $n$ of identical partly entangled pairs, one can concentrate
their entanglement into a smaller number of maximally entangled pairs
such as singlets.  This process of ``entanglement concentration'' is
asymptotically efficient in the sense that, for large $n$, the yield
of singlets approaches $nE-O(\log n)$.  Conversely, local operations can be used
to prepare arbitrary partly-entangled states $\Psi(AB)$ of two
subsystems from a starting material consisting of standard
singlets, again in a manner which asymptotically conserves entropy
of entanglement.
 
We should clarify what we mean by local operations.  Initially
the $n$ partly-entangled pairs are shared between two parties
(call them Alice and Bob) with Alice receiving one member of each pair,
and Bob receiving the other.  This
non-local sharing establishes an initial entanglement
$nE$ between Alice and Bob.  After that Alice and Bob operate
locally on their particles, with Alice for example performing
unitary operations and von Neumann or generalized measurements in
the Hilbert space of her particles, and Bob
performing similar operations in that of his particles.
We allow Alice and Bob to coordinate their actions through
exchange of classical messages, but not to exchange
any quantum systems nor to perform any nonlocal operation after
the initial sharing.  This restriction is of course necessary to
force Alice and Bob to use the partly entangled pairs they
already have, rather than generating perfectly entangled pairs de
novo.

\section{Entanglement Concentration}
In this section we describe a method whereby the entanglement present in a
supply of identical partially-entangled pairs of 2-state particles can be
concentrated into a smaller number of perfect singlets.  The generalization
to $d>2$ state particles is straightforward.  We call the method Schmidt
projection because its essential step is a projection of the joint state of
$n$ pairs of particles onto a subspace spanned by states having a common Schmidt
coefficient.
 
Let $n$ partly-entangled pairs of 2-state particles be shared between Alice
and Bob, so that the initial state is
 
\beq
\Psi(A,B) = \prod_{i=1}^n(\cos\theta\ket{\alpha_1(i)\beta_1(i)}
                   +\sin\theta\ket{\alpha_2(i)\beta_2(i)}).
\eeq
When binomially expanded, this state has $2^n$ terms, with only
$n+1$ distinct coefficients, $\cos^n\theta, \cos^{n-1}\theta \sin
\theta...\sin^n\theta$.  Let one of the parties (say Alice)
perform an incomplete von Neumann measurement projecting the
initial state into one of $n+1$ orthogonal subspaces
corresponding to the power $k=0...n$ to which $\sin\theta$
appears in the coefficient.  Either party can perform this
measurement locally, Alice by measuring the particles she has,
or Bob by measuring the ones he has.  Let Alice perform
the measurement, obtaining some outcome $k$.  She then tells Bob
which outcome she obtained.  Alternatively, if Bob and Alice
wish not to communicate, Bob can perform his version of the
measurement locally, and, by virtue of the original
entanglement, he will always obtain the same value of $k$ as
Alice has.   The probability of outcomes is binomially
distributed, with outcome $k$ having probability
\beq
p_k= \choose{n}{k}(\cos^2\theta)^{n-k}(\sin^2\theta)^k.
\eeq
 
After some outcome $k$ has been obtained, Alice and Bob will be left
with a residual state $\Psi_k$ of their spins which is a maximally
entangled state in a known $2\choose{n}{k}$-dimensional subspace of the
original $2^{2n}$ dimensional space.  Such states can be used without
further ado for faithful teleportation in an $\choose{n}{k}$-dimensional
or smaller Hilbert space; or they can be transformed, as described
below, into a standard form such as singlets.
 
Before describing this optional standardization process, we note that
the measurement of $k$ occasionally yields a residual state $\Psi_k$
with {\em more\/} entropy of entanglement than the original state
$\Psi$.  However, neither the measurement of $k$ nor any other local
processing by one or both parties can increase the {\em expected\/}
entropy of entanglement between Alice's and Bob's subsystems.  Consider
a measurement or other local treatment applied by Alice, resulting in a
classical outcome $j$ and a residual pure state $\Psi_j$ of the joint
system.  This treatement cannot influence the partial density matrix
$\rho_B$ seen by Bob, since if it did one would have a superluminal
communications channel based on Alice's applying or not applying the
treatment and Bob measuring $\rho_B$. Therefore, depending on the extent
of correlation between the residual state and the classical outcomes,
the expected entanglement of the residual states lies between $E(\Psi) -
H$ and $E(\Psi)$, where $E(\Psi)$ is the original pure state's
entanglement and $H=-\sum_jp_j\log_2p_j$ is the Shannon entropy of the
measurement outcomes.  All local treatments (eg generalized or positive
operator valued measurements~\cite{Helstrom}) that Alice might apply can be cast
in this form, if
necessary by considering her operations to be performed in an
appropriately enlarged Hilbert space.  In particular, unitary
transformations by Alice correspond to one-outcome measurements, which
cannot change Bob's partial density matrix $\rho_B$ at all, and can only
change the eigenvectors, but not the eigenvalues, of Alice's.  By the
same argument, local actions by Bob cannot increase the expected
entanglement between his and Alice's subsystems.

Thus, though Alice and Bob cannot by local actions increase their
expected entanglement, they can gamble with it, spending their initial
amount on a chance of obtaining a greater amount.

We now show how the entanglement in the above perfectly-entangled
residual states $\Psi_k$ can be efficiently transformed into a standard
form such as singlets.  Fix some small positive $\epsilon$, with
$\epsilon=0$ corresponding to perfect efficiency of transformation. Let
the above measurement of $k$ be performed independently on a sequence of
batches of $n$ pairs each.  Each performance yields another $k$ value;
let the resulting sequence of $k$ values be $k_1, k_2, ...k_m$, and let
 \beq
D_m=\choose{n}{k_1}\choose{n}{k_2}...\choose{n}{k_m}
 \eeq be the product of the $\choose {n}{k}$ values for the first $m$
batches.  The sequence is continued until the accumulated product $D_m$
slightly lies between $2^\ell$ and $2^\ell(1+\epsilon)$ for some power
$\ell$.  For any single-pair entanglement $E$ and any positive
$\epsilon$, the probability of failing to come this close to a power of
two tends to zero with increasing $m$.  Once a suitable $D_m$ is found,
a local measurement is performed by Alice or Bob or both to project the
joint system into one of two orthogonal subspaces, a large space of
dimension $2\cdot2^\ell$ and a relatively small space of dimension
$2\cdot(D_m-2^\ell)<\epsilon\cdot2\cdot2^\ell$.  In the latter case,
occurring with probability less than $\epsilon$, a failure has occurred,
and all or most of the entanglement will have been lost.  In the former
case, occurring with probability greater than $1-\epsilon$, the residual
state is a maximally entangled state of two $2^l$ dimensional
subsystems, one held by Alice and one held by Bob.  Using the Schmidt
decomposition, this can be converted by local unitary operations into a
product of $\ell$ standard singlets.

\section{Efficiency}
We deal first with the efficiency of the initial concentration
stage, which yields a binomially distributed measurement result
$k$ and collapses the initial state of $n$ partly-entangled pairs
into a maximally-entangled state between two systems of
dimensionality $\choose{n}{k}$.  We show below that its
efficiency approaches unity for large $n$.  Then we show that the
second, or standardization, stage, which distills standard
singlets from these high-dimensional maximally-entangled states,
also approaches unit efficiency for large $m$.
 
We adopt the local viewpoint of one of the observers, say Alice.
From her viewpoint, the initial state is a mixed state of entropy
$nE$.  Performing the measurement of $k$ splits off some of this
entropy in the form of the entropy of the distribution of
outcomes $k$, and leaves the rest of it as concentrated entropy
of entanglement between the two residual maximally-entangled
$\choose{n}{k}$-dimensional subsystems.  The expected amount of
concentrated entropy of entanglement is given by
\beq
\sum_{k=1}^{n-1} (\cos^2\theta)^{n-k} (\sin^2\theta)^k \,\,
\choose{n}{k} \log_2\choose{n}{k}.
\eeq
Because the entropy of the binomial distribution of $k$ values
increases only logarithmically with $n$, the fraction of the
original entanglement $nE$ captured as concentrated
entanglement approaches 1 in the limit of large $n$.
 
We now the show that the efficiency of distilling standard
singlets from large-dimensional fully-entangled
states produced by the measurements of $k$ also tends to unity.
In the light of the previous discussion, it suffices to show
that for any fixed batch size $n>1$, the sequence
\beq
Z_m = \mod(\log_2 D_m,1) = \log_2 D_m -\ell
\eeq
of mantissas of base-2 logarithms of $D_m$, where $D_m$ is given
in equation 7, has an infimum of 0.  This in turn follows from
the fact that the binomial coefficients
$\choose{n}{k_1},\choose{n}{k_2}...$ are independently drawn from
a fixed distribution (eq 6).  The evolution of $Z_m$ with
increasing $m$ may therefore be viewed as a random walk on the
unit interval (with wraparound), starting at the origin and
taking steps of sizes $\; \mod(\log_2\choose{n}{k},1) \:$ and
probabilities given by equation 6.  It is elementary to show that
for any distribution of step sizes, and any positive $\epsilon$,
such a walk visits the interval $[0,\log_2(1+\epsilon)]$ with
probability 1, from which the theorem follows.
 
The Schmidt projection method of entanglement concentration requires at
least $n=2$ partly-entangled pairs, and only becomes efficient for large
$n$.  We now describe another method that works, albeit inefficiently,
even with a single partly entangled pair, as in eq.~1.  We call this
procedure the Procrustean method of entanglement concentration, because
its goal is to cut off and discard the extra probability of the larger
term in equation 1, leaving a perfectly entangled state.  Assume for the
moment that $\theta < \pi/4$ so that if Alice measures particle $1$ in
the up/down basis, the up outcome is more likely.  Instead of performing
this von Neumann measurement, she passes her particle through a
polarization-dependent absorber, or a polarization-dependent-reflector
(eg, for light, a Brewster window), which has no effect on down spins
but absorbs, or deflects into a different beam, a fraction
$\tan^2\theta$ of the up spins.  If the particle is absorbed or
deflected, it is rejected; otherwise it is kept. This treatment does not
correspond to any von Neumann measurement in the original 2-dimensional
spin space, but rather to a two-outcome generalized measurement or POVM
(positive-operator-valued measurement)~\cite{Helstrom,Gisin}. If the
particle is not absorbed or deflected, its residual state after this
treatment will be a maximally mixed state of spin up and spin down.  Now
suppose Alice tells Bob the result of her generalized measurement, and
suppose that he does not measure his particle at all, but simply
discards it if Alice has discarded hers.  The result will be a perfectly
entangled state of two particles.  The Procrustean method is especially
suitable for the type of gambling mentioned earlier: when it works,
it always yields more entanglement than the parties started out with.
 
Both the Schmidt projection and the Procrustean method
can be generalized to work on larger Hilbert spaces.
Like von Neumann's method for obtaining unbiased random bits from
a coin of unknown but time-independent head/tail ratio
(setting HT=1, TH=0, and TT=HH=\mbox{do over}),
Schmidt projection works even when Alice and Bob do not know how
entangled their partly-entangled pairs are, provided all $n$
pairs have equal biases $\theta$.  The Procrustean method, on the
other hand, requires the bias to be known in advance.
 
Figure 1 plots the yield of perfectly entangled pairs
as a function of $\cos^2\theta$ obtained by the Schmidt projection method
with $n=2,4,8,$ and $32$ (lower 4 curves), in comparison with
ideal asymptotic yield $nE=nH_2(\cos^2\theta)$, (top curve) and the
yield from the Procrustean method (inverted-V shaped curve).
Note that for $n<5$ Schmidt projection is absolutely less
efficient than the Procrustean method.

\section{Relation of Entanglement Concentration to Quantum
Data Compression}
 
Like the Schmidt projection method of entanglement concentration
described above, the technique of {\sl quantum data
compression\/}~\cite{JS94,Schu95} involves projecting the state of a
high-dimensional system onto a set of orthogonal subspaces depending on
eigenvectors and eigenvalues of an associated density matrix.  However,
the goals and means of the two techniques are sufficientlly different
that neither can be substituted for the other. Indeed, certain quantum
data transmission tasks can only be accomplished efficiently by using
the two techniques together.
 
Quantum data compression (QDC) has the goal of encoding an unknown
sequence of signals from a known quantum {\sl source\/}---ie an ensemble
of pure states $\{\psi_j\}$ emitted with specified probabilities
$\{p_j\}$---into a smaller Hilbert space than it originally occupies,
while introducing negligible distortion.  QDC is useful when the source
has less than maximal entropy (permitting it to be compressed at all) and
consists of non-orthogonal states $\psi_j$, necessitating the use of
quantum operations to do the compression.
 
As has previously been noted~\cite{Schu95,BBJMPSW94,JSV95}, a
quantum  source is not fully specified by its density matrix
$\rho = \sum_jp_j \proj{\psi_j}$.  By the same token, it is also not
fully specified by giving an entangled state $\Psi_{AB}$ of which its
density matrix is the partial trace, eg
$\rho=\rho_A=\Tr_B\proj{\Psi_{AB}}$.  A quantum source can, however, be
fully specified by giving both such an entangled state $\Psi_{AB}$ and a
von Neumann or generalized measurement to be performed by Bob, who holds
subsystem $B$.  This is done in such a way that each of Bob's possible
measurement outcomes projects Alice's subsystem into one of the states
$\psi_j$, and the outcomes occur with the required probabilities
$\{p_j\}$. Then each of Bob's measurements tells him which state Alice
received from the source at that instant.  For example, depending on the
measurement performed by Bob, the entangled state of equation 1 can be
used to generate either of the following two sources for Alice, one
classical, the other distinctively quantum.
 \begin{itemize}
\item Source $Q$, consisting of orthogonal states
$\ket{\uparrow}$ and
$\ket{\downarrow}$ emitted with unequal probabilities \,
$\cos^2\theta$\, and \, $\sin^2\theta$,\, respectively; and
\item Source $Q'$ consisting of non-orthogonal states
$\psi^0=(\cos\theta\ket{\uparrow}+\sin\theta\ket{\downarrow})$
 and
$\psi^1=(\cos\theta\ket{\uparrow}-\sin\theta\ket{\downarrow})$
emitted with equal probabilities.
\end{itemize}
 
The first source, $Q$, is purely classical in the sense that it could
be faithfully compressed by making a complete von Neumann
measurement in the up/down basis, and applying conventional data
compression (eg Huffman coding) to the resulting classical data.
 
Although the other source, $Q'$, would yield statistically
similar data when measured in the up/down basis, the resulting
data would be useless for reliably encoding a sequence of $n$
states from the source, because the data would
be utterly uncorrelated with which of the $2^n$ equiprobable
non-orthogonal spin sequences the source had emitted.
 
At this point some notation is helpful.  Let $x$ denote an arbitrary
$n$-bit sequence, where $n>2$, and let $\Psi^x$ denote the $n$-spin
product state resulting when source $Q'$ emits a sequence of states
indexed by the bits of $x$.  For example, taking $x=011$, we have
$\Psi^{011} = \psi^0 \otimes \psi^1 \otimes \psi^1$.
 
In order to transmit such sequences faithfully and economically,
one uses quantum data compression~\cite{JS94,Schu95}.  This consists of
performing a very gentle, incomplete measurement on the the joint state
$\Psi^x$ of the spins, which projects the state into one of two complementary
subspaces :
 \begin{itemize}
 \item a ``likely'' subspace of
dimensionality $2^{n(H(\rho)+\delta)}$ spanned by the eigenvectors of
the largest eigenvalues of the joint density matrix $\rho^{(n)}$, defined
as the tensor product of $n$ copies of $\rho$.
 \item the ``unlikely'' subspace spanned by the remaining eigenvectors.
\end{itemize}\noindent
If the joint state projects into the likely
subspace, one transmits the resulting projected state; if the
projection fails, one transmits an arbitrary state.  Using $W^x$ to
denote the (slightly mixed) state resulting from applying quantum data
compression to a source sequence $\Psi^x$ from the source $Q'$, it
can be shown that  for any positive $\delta$ and $\epsilon$ there exists
an $n_0$ such that for all $n>n_0$, the fidelity of the quantum
data compression, ie the probability
 \beq
 F = \sum_x p(x) \bra{\Psi^x}W^x\ket{\Psi^x}
 \eeq that its output would pass a test for being the same as the input
sequence $\Psi^x$, conducted by someone who knew what the input sequence
was, is greater than $1-\epsilon$. Infidelity can be thought of as
resulting from two causes: failure to project into the likely subspace,
and failure of even a successful projection into that subspace to agree
with the original state when subsequently tested.  Both kinds of
infidelity become negligible in the limit of large
$n$~\footnote{Although one is usually more interested in data
compression, it is also possible to perform classical or quantum data
{\em expansion\/}, in other words to encode the output of a nonredundant
source into a larger number of redundant bits or qubits. For example,
all states of two qubits can be encoded into the majority-up subspace
of three qubits, each of which has a single-particle density matrix with
eigenvalues 3/4 and 1/4 and entropy of approximately 0.811 bits.}.
 
Because a less-than-maximally entangled pure state appears
as a less-than-maximally random mixed state to its two separate
observers, one could imagine using quantum data compression
as an approximate means of entanglement concentration.  In other
words, by separately compressing their respective subsystems, Alice and
Bob could squeeze the original entanglement into a smaller number
of shared pairs of qubits.  Applying this two-sided compression to
$n$ shared pairs of entanglement $E$, Alice's and Bob's projections
into the likely subspace would be isomorphic (this can be seen by
considering the Schmidt representation of the original entangled
state), and the result would be to leave each observer with
slightly more than $nE$ qubits having slightly less than $nE$ bits
entropy of entanglement with an equal number of qubits held by
the other observer.  However, the entangled sates produced by such
two-sided quantum data compression are never maximally entangled, and in
an important sense they become {\sl poorer\/} approximations of
maximally-entangled states as $n$ increases.  A perfectly-entangled
state must have all its Schmidt coefficients equal, but the compressed bipartite
state $\Psi_c(A,B)$, resulting from applying two-sided quantum data compression
to a state such as the $\Psi(A,B)$ of eq.~5, describing a set of $n$
partly-entangled pairs, has
a distribution of values of Schmidt coefficents whose variance increases
with $n$.  In consequence, the fidelity with which $\Psi_c(A,B)$
approximates any maximally entangled state approaches zero with
increasing $n$, as does the fidelity with which it would work in
teleporting a random state in a Hilbert space of dimensionality equal to
that of one of its two parts.

We have just shown that two-sided quantum data compression does not work
as a method of entanglement conentration. Conversely, because any mixed
state can be regarded as the partial trace of an appropriate entangled
state, one can imagine attempting to use the Schmidt projection method
of entanglement concentration in a one-sided manner as a way of
performing quantum data compression.  As we will show presently, this
also does not work. To use Schmidt projection in a one-sided manner
would mean making a more aggressive projection than that used in
conventional quantum data compression, into subspaces spanned by each
distinct eigenvalue of $\rho^{(n)}$,
rather than into a single subspace spanned by all the likely eigenvalues
and a residual unlikely subspace.  However, this projection is too
aggressive for the purposes of reliable data transmission.  Because the
entropy of the distribution of the eigenvalues increases absolutely with $n$,
(although it decreases as a fraction of $n$), the fidelity of
transmission of typical sequences from a source such as $Q'$ tends to
zero with increasing $n$.

In more detail, the proposed Schmidt projection method of quantum data-compression, whose
fidelity we seek to refute, corresponds in the case of the
source $Q'$ to an incomplete
measurement, in which one observes the number $k$ of down
spins in an $n$-spin block, leaving a residual quantum state
equal to the renormalized projection of the original source
sequence into the subspace corresponding to the measurement
outcome.  We shall show that this coding {\em cannot\/}
transmit sequences from the source $Q'$ with asymptotically
perfect fidelity, even if the measurement outcome $k$ is made
available, as classical information, to help in the decoding
process at the receiving end of the channel.  Such a dual
classical/quantum channel setup is certainly no weaker than a
purely quantum channel (where the classical outcome $k$ would
simply be discarded) and is analogous to the dual channel used in
teleportation, in which the quantum channel output is
postprocessed in a way depending on classical information
generated at the sending end.  The inability of even this strong
kind of Schmidt coding to approach perfect fidelity for source
$Q'$ follows from the fact, to be demonstrated presently, that
with significant probability it maps distinct input sequences
onto the same output state.
 
Recalling notation introduced earlier,  we use $\Psi^x$ to denote
the $n$-spin product state resulting when source $Q'$ emits a
sequence of states indexed by the bits of bit string $x$.  For example,
taking $x=011$, we have $\Psi^{011} = \psi^0 \otimes \psi^1
\otimes \psi^1$.  Let $\bar{x}$ denote the Boolean complement of
$x$.  In our example $\bar{x}$ would be $100$, and
correspondingly $\Psi^{\bar{x}}$ would be $\Psi^{100} = \psi^1
\otimes \psi^0 \otimes \psi^0$.  By expanding the states $\Psi^x$
and $\Psi^{\bar{x}}$ in the up/down basis, and grouping terms
according to $k$, it can readily be seen that the states $\Psi^x$
and $\Psi^{\bar{x}}$ differ only by a uniform sign change in all
terms of odd $k$.  This difference is obliterated by the Schmidt
encoding process, which by measuring $k$ randomizes the relative
phases of terms of differing $k$.  Thus, for any $x$, the
encoding process, even if the classical outcome $k$
is made available at the receiving end, maps input states
$\Psi^x$ and $\Psi^{\bar{x}}$ onto the same output state or
distribution of output states.  This precludes the achievement of
asymptotically perfect fidelity, which by definition requires
that with probability approaching unity a typical channel input
be mapped onto a pure or mixed output state arbitrarily close to
the input in Hilbert space.  Here, on the contrary, we have
distinct inputs $\Psi^x$ and $\Psi^{\bar{x}}$, which for large $n$
are nearly orthogonal, being mapped onto the same output.
 
In summary, two-sided quantum data compression is too faithful to the
original non-maximally-entangled state to provide good entanglement
concentration (whose goal is a maximally entangled state rather than a
good approximation to the initial state).  Conversely, entanglement
concentration by the Schmidt projection method necessarily sacrifices
fidelity to the original state in order to produce a maximally entangled
output.
 
Despite these differences, the two techniques can sometimes be fruitfully
combined.  Suppose Alice has a long sequence of $n$ spins from a
nonrandom quantum source such as $Q'$ which she wishes to faithfully
teleport to Bob with a minimal usage of
engtanglement.  Suppose further that the entanglement Alice and Bob
have at their disposal for teleportation purposes is supplied not
in the form of standard singlets but rather as pure but less-than-maximally
entangled states $\psi_{AB}$.  An economical procedure would then be
 \begin{itemize}
 \item use quantum data compression to compress the
input sequence to a bulk of slightly greater than $nH(\rho(Q'))$ spins.
\item use entanglement concentration to prepare standard singlets
from the supply of imperfectly entangled pairs.
 \item teleport each compressed spin using one of the standard singlets.
\end{itemize}

\section{Discussion}

We have shown that the entanglement in any pure state of a bipartite
system can be concentrated by local operations and classical
communication into maximally entangled states such as singlets. Here we
note that, conversely, an arbitrary partly-entangled state $\Psi(A,B)$
of a bipartite system can be prepared by local operations and classical
communication using standard singlets as the only source of
entanglement. One way to do this is for Alice first to prepare a copy
$\Psi(A,C)$ of the entangled state she wishes to share with Bob, using
two systems in her laboratory, $A$ and $C$.  Here $A$ is the system she
wishes to entangle with Bob's system $B$, and $C$ is a system similar to
$B$ but located in Alice's laboratory instead of Bob's.  Because $A$ and
$C$ are in the same location, $\Psi(A,C)$ can be prepared by purely
local operations.  Next Alice uses a supply of standard singlets, in
conjunction with classical communication, to teleport the state of her
local system $C$ into Bob's system $B$. This has the
effect~\cite{BBCJPW93} of destroying the entanglement between $A$ and
$C$ while creating the desired entangled state $\Psi(A,B)$ shared
remotely between Alice and Bob.  The teleportation consumes $\log_2d$
singlets and requires $2\log_2d$ classical bits of communication, where
$d$ is the dimension of the Hilbert space of system $B$, regardless of
the entanglement $E$ of the state being teleported, which is rather
inefficient if $E$ is small. To reduce the consumption of singlets
toward the theoretical minimum, Alice need only use Schumacher coding to
compress her $C$ systems before teleporting them to Bob, who then
applies the inverse decoding operation to re-expand the teleported $B$
systems to their original dimensions.
 
Because entropy of entanglement can be thus efficiently concentrated and
diluted, its positive-definiteness and the nonincrease of its expecation
under local operations represent the {\em only\/} limitation on the
asymptotic ability of two separated observers to prepare a final pure
state $\Psi'_{AB}$ given an inital one, $\Psi_{AB}$, by local operations
and classical communication.  Thus if $\Psi$ and $\Psi'$ have equal
entanglement, they can be interconverted with efficiency approaching
unity in the limit of large $n$; if they have unequal entanglement, the
asymptotic yield is $E(\Psi')/E(\Psi)$.  On the other hand, if the
initial state is unentangled, the positive-definiteness of entanglement
and the nonincrease of its expectation under local operations together
imply that there is no way of locally preparing an entangled final state
$\Psi'$, even with low yield.
 
We have argued that entropy of entanglement is a good entanglement
measure for pure states because local operations cannot increase
its expectation, but can, with asympotically perfect efficiency,
interconvert states of equal entropy of entanglement.  No similarly
simple entanglement measure is known for mixed states.  For a mixed
state, the subsystem entropies $H_A$ and $H_B$ need not be equal, and
neither they nor any other simple function of $H_A$, $H_B$, and the
total entropy $H_{AB}$ can be clearly identified with the degree of
entanglement.  For example, entropies of $H_A=H_B=H_{AB}=2$ can be
realized by two very different mixed states:
 \begin{itemize}
 \item two independent pairs of classically correlated spins, or
 \item one singlet and a pair of uncorrelated spins.
 \end{itemize} The former mixed state is naturally regarded as
unentangled, since it can be made by mixing unentangled pure states; the
latter should be regarded as entangled, since it can be converted into a
maximally-entangled state by discarding the second pair of spins.

Two possible measures of entanglement for the mixed
bipartite state $\rho_{AB}$, each of which reduces to
entropy of entanglement when $\rho_{AB}$ is pure, are:
 \begin{itemize}
 \item ``Entanglement of formation'' defined as the least
number of shared singlets asymptotically required to
prepare $\rho_{AB}$ by local operations and classical communication.
 \item ``Distillable entanglement'' defined as the greatest
number of pure singlets that can asymptotically be prepared
from $\rho_{AB}$ by local operations and classical communication.
 \end{itemize}
These entanglement measures have the desirable feature that
their expectations
cannot be increased by local operations, but the disadvantage of being
hard to evaluate in particular cases because of the
implied optimizations over local procedures.

Since any mixed state can be regarded as a pure state over a
larger, and partly unseen, Hilbert space, it might seem more
elegant (cf~\cite{J94}) to define the entanglement of a bipartite 
mixed state $\rho_{AB}$ as the minimum entanglement of any pure state
$\Psi(ABCD)$ of a four-part system, from which $\rho_{AB}$ can
be obtained by tracing over states of $C$ and $D$, and where the
entanglement of $\Psi(ABCD)$ is computed by partitioning it into
subsystems $AC$ and $BD$.  Unfortunately, this prescription
is unsatisfactory because it would assign positive
entanglement to classically correlated mixed states such as $\rho
= \half (\proj{\uparrow\uparrow} + \proj{\downarrow\downarrow})$,
which can be generated from unentangled initial conditions by
local actions (random choice) and exchange of classical messages.

A nontrivial example of an entangled mixed state is provided by
the Werner state~\cite{Werner}, whose density matrix is
 \beq
W = \frac{1}{8}I + \frac{1}{2}\proj{\Psi^-}
 \eeq where $I$ is the 4x4 identity matrix and $\Psi^-$ is a singlet.
This state may be viewed as a 50/50 mixture of totally mixed states
(density matrix $\frac{1}{4}I$) and singlets; and indeed, since totally
mixed states can be manufactured by purely local means, this description
constitutes a local probabilistic procedure for generating $n$ Werner
states from an expected $n/2$ singlets.  Recently it has been
shown~\cite{BBPSSW95}
that the Werner state can be constructed much more economically, as a
mixture of pure states each containing only about 0.1176 ebits of
entanglement.  Moreover, a small yield $\geq 0.0001258$ of arbitrarily
pure singlets can be distilled from Werner states by local actions and
classical communication~\cite{BBPSSW95}.  Thus impure Werner states can
be both created from and converted into pure singlets.  This two-way
conversion is a rough mixed-state analog of the processes of pure-state
entanglement concentration and dilution which have been the principal
subject of the present paper.  However, it is not known whether the
interconversion is asymptotically reversible for mixed states, as it is
for pure states. In other words, it is not known whether the Werner
state's distillable entanglement is equal to, or far less than, its
entanglement of formation.

\section*{Acknowledgments} We wish to thank William Wootters for advice
on Schmidt decompositions and many other subjects, and John Smolin for
pointing out how to use teleportation to prepare arbitrary entangled
states.

\newpage
\section*{Figure Caption}
\begin{enumerate}
\item Yield of maximally-entangled output states from partly-entangled input
states $(\cos\theta \uparrow\downarrow-\sin\theta\downarrow\uparrow)$,
as a function of $\cos^2\theta$.
Highest curve is entropy of entanglement of input state,
equal to the asymptotic yield of the Schmidt projection method.
Successively lower smooth curves give yields of Schmidt projection
applied to $n=$ 32, 8, 4, and 2 input pairs.
The $\Lambda$-shaped curve gives yield by the Procrustean method
applied to one input pair.
\end{enumerate}

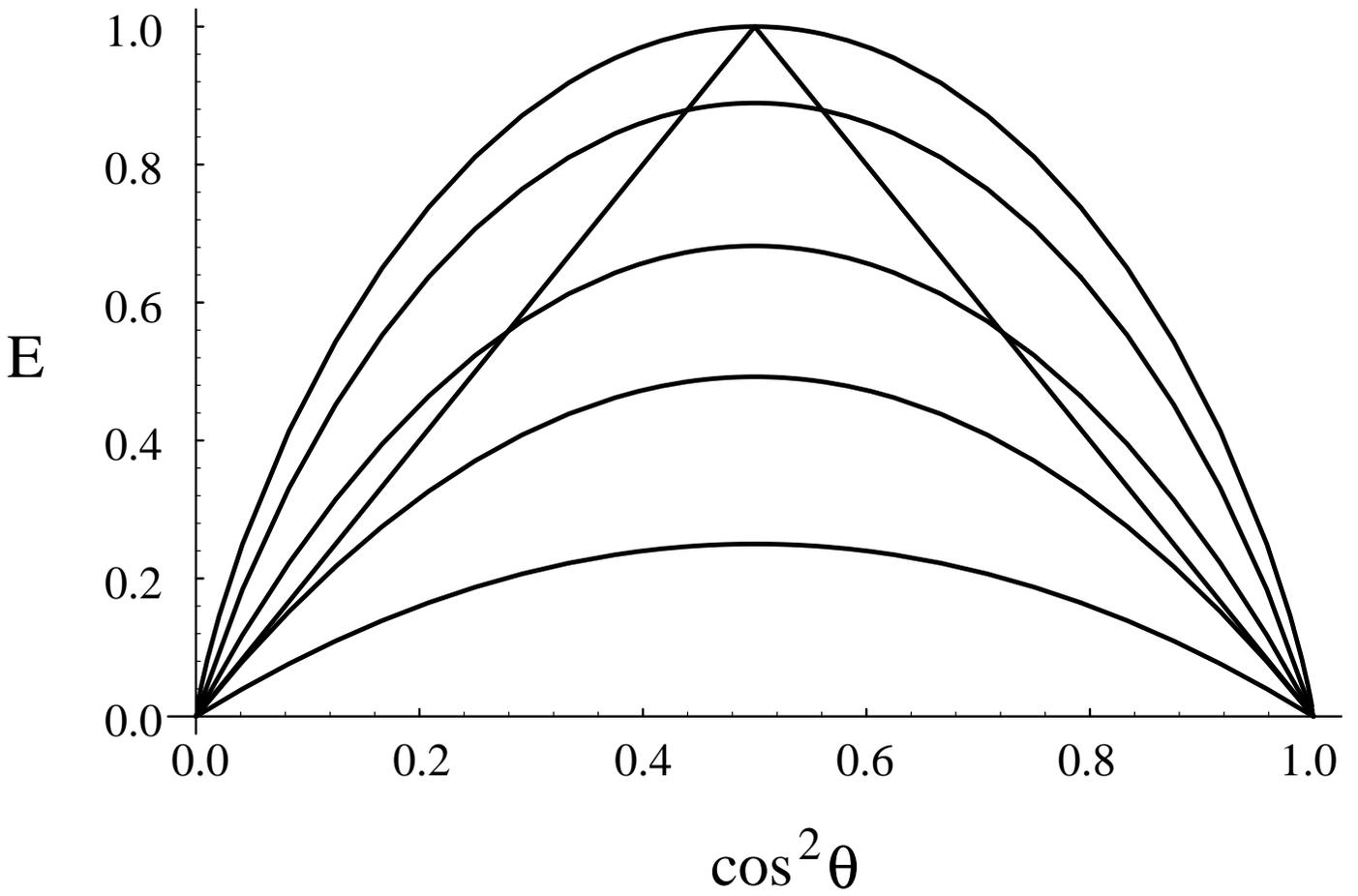
\end{document}